\documentclass[fleqn,usenatbib]{mnras}

\usepackage{newtxtext,newtxmath}

\usepackage[T1]{fontenc}
\usepackage{ae,aecompl}

\usepackage{graphicx}	
\usepackage{amsmath}	
\usepackage{amssymb}	
\usepackage{color}
\usepackage{algorithm}
\usepackage[noend]{algpseudocode}
\usepackage{pifont}
\usepackage{soul}
\usepackage{float}
\usepackage{multibib}
\newcites{Appendix}{Appendix References}
\usepackage{natbib}
\usepackage[flushleft]{threeparttable}
\usepackage{float}
\usepackage{placeins}
\usepackage{lscape}
\usepackage{lipsum}
\usepackage{array}
\usepackage{flafter}
\usepackage{bookmark}

\graphicspath{{Figures/}}

\newcommand*\diff{\mathop{}\!\mathrm{d}}
\DeclareMathOperator\erf{erf}
\DeclareMathOperator\erfc{erfc}

\title[From 2D projections to 3D fractal cloud structures]{The relation between the true and observed fractal dimensions of turbulent clouds}  

\author[J. R. Beattie, C. Federrath and R. S. Klessen]{
James R. Beattie$^{1,2}$\thanks{E-mail: beattijr@mso.anu.edu.au}, Christoph Federrath$^{1}$\thanks{E-mail: christoph.federrath@anu.edu.au} and Ralf S. Klessen$^{3,4}$
\\
$^{1}$Research School of Astronomy and Astrophysics, Australian National University, Canberra, ACT 2611, Australia\\
$^{2}$Science and Engineering Faculty, Queensland University of Technology, Brisbane, QLD 4000, Australia\\
$^{3}$Universit\"at Heidelberg, Zentrum f\"ur Astronomie, Institut f\"ur Theoretische Astrophysik, Albert-Ueberle-Str. 2, 69120 Heidelberg, Germany\\ 
$^{4}$Universit\"at Heidelberg, Interdisziplin\"ares Zentrum f\"ur Wissenschaftliches Rechnen, Im Neuenheimer Feld 205, 69120 Heidelberg, Germany
}

\date{Accepted XXX. Received YYY; in original form ZZZ}

\pubyear{2019}

\begin{document}
\label{firstpage}
\pagerange{\pageref{firstpage}--\pageref{lastpage}}
\maketitle

\begin{abstract}
Observations of interstellar gas clouds are typically limited to two-dimensional (2D) projections of the intrinsically three-dimensional (3D) structure of the clouds. In this study, we present a novel method for relating the 2D projected fractal dimension ($\mathcal{D}_{\text{p}}$) to the 3D fractal dimension ($\mathcal{D}_{\text{3D}}$) of turbulent clouds. We do this by computing the fractal dimension of clouds over two orders of magnitude in turbulent Mach number $(\mathcal{M} = 1-100)$, corresponding to seven orders of magnitude in spatial scales within the clouds. This provides us with the data to create a new empirical relation between $\mathcal{D}_{\text{p}}$ and $\mathcal{D}_{\text{3D}}$. The proposed relation is \mbox{$\mathcal{D}_{\text{3D}}(\mathcal{D}_{\text{p}}) = \Omega_1 \erfc ( \xi_1 \erfc^{-1}[ (\mathcal{D}_{\text{p}} - \mathcal{D}_{\text{p,min}})/\Omega_2 ] + \xi_2 ) + \mathcal{D}_{\text{3D,min}}$}, where the minimum 3D fractal dimension, $\mathcal{D}_{\text{3D,min}} = 2.06 \pm 0.35$, the minimum projected fractal dimension, $\mathcal{D}_{\text{p,min}} = 1.55 \pm 0.13$, $\Omega_1 = 0.47 \pm 0.18$, $\Omega_2 = 0.22 \pm 0.07$, $\xi_1 = 0.80 \pm 0.18$ and $\xi_2 = 0.26 \pm 0.19$. The minimum 3D fractal dimension, $\mathcal{D}_{\text{3D,min}} = 2.06 \pm 0.35$, indicates that in the high $\mathcal{M}$ limit the 3D clouds are dominated by planar shocks. The relation between $\mathcal{D}_{\text{p}}$ and $\mathcal{D}_{\text{3D}}$ of molecular clouds may be a useful tool for those who are seeking to understand the 3D structures of molecular clouds, purely based upon 2D projected data and shows promise for relating the physics of the turbulent clouds to the fractal dimension.    
\end{abstract}


\section{Introduction}\label{introduction}
The aim of this study is to introduce a new method for calculating the fractal dimension of turbulent clouds, and to provide an empirical relation from the fractal dimension of a 2D projection of the cloud to the true 3D fractal dimension. Creating such a relation elucidates some geometrical properties of the 3D structure of interstellar clouds, based purely upon information from the 2D projection. 


The structure of molecular clouds in the interstellar medium (ISM) is of great interest. For example, filaments in molecular clouds (MCs) give rise to local, high-density regions that provide the necessary conditions for the action of gravity to collapse parts of the cloud into what may eventually become a star  \citep{Scalo1997,Ferriere2001,MacLow2004,Kainulainen2009,Arzoumanian2011,Federrath2012,Federrath2013,Andre2014,Federrath2016,Hacar2018,Mocz2018}. 

Most observations of MCs are 2D projections of the 3D clouds and information about the 3D volumetric properties of the cloud in position-position-position (PPP) is mostly lost and highly challenging to reconstruct \citep{Larson1981,Brunt2010b,Brunt2010a,Beaumont2013,Ginsburg2013,Brunt2014,Tritsis2018}. In this paper, we construct an empirical relation from the 2D cloud projection (in PP, which we call ``the observed cloud") to the 3D cloud (in PPP, which we call the ``true" cloud), based on the fractal dimension. As far as we know, no such relation has been directly measured from turbulent MCs, and only simple analytical forms developed for ideal fractals have been explored in the past (see Equation 5 in \citealt{Sanchez2005}, for example). Such a relation will provide insight into how real clouds are organised by quantifying the amount of diffuse, filamentary or sheet-like structures of MCs in the ISM. 

In \S \ref{sec:theory} of this study we outline the required turbulence and fractal geometry theory. In \S \ref{sec:sims} we give details on the extensive suite of simulations that we analyse, and the data types that we use to explore the turbulent clouds. In \S \ref{sec:MethodML} we describe a new method that we use to calculate mass-length and fractal dimension curves from the simulated clouds. Following, in \S \ref{sec:MethodCompCurves} we put the curves into a common frame of reference, creating composite fractal dimension curves. We then use \cite{Mandelbrot1983}'s relation to map the 2D slice data to the 3D data and construct empirical fits for the 3D and 2D fractal dimension curves. Next, in \S \ref{sec:relation} we introduce the new relation from the 2D projections (PP) to the 3D cloud (PPP) data, and suggest how this may be used in astrophysical applications. Finally, in \S \ref{sec:conclusions} we summarise our key findings.

\begin{figure*}
\centering
\includegraphics[width=0.95\linewidth]{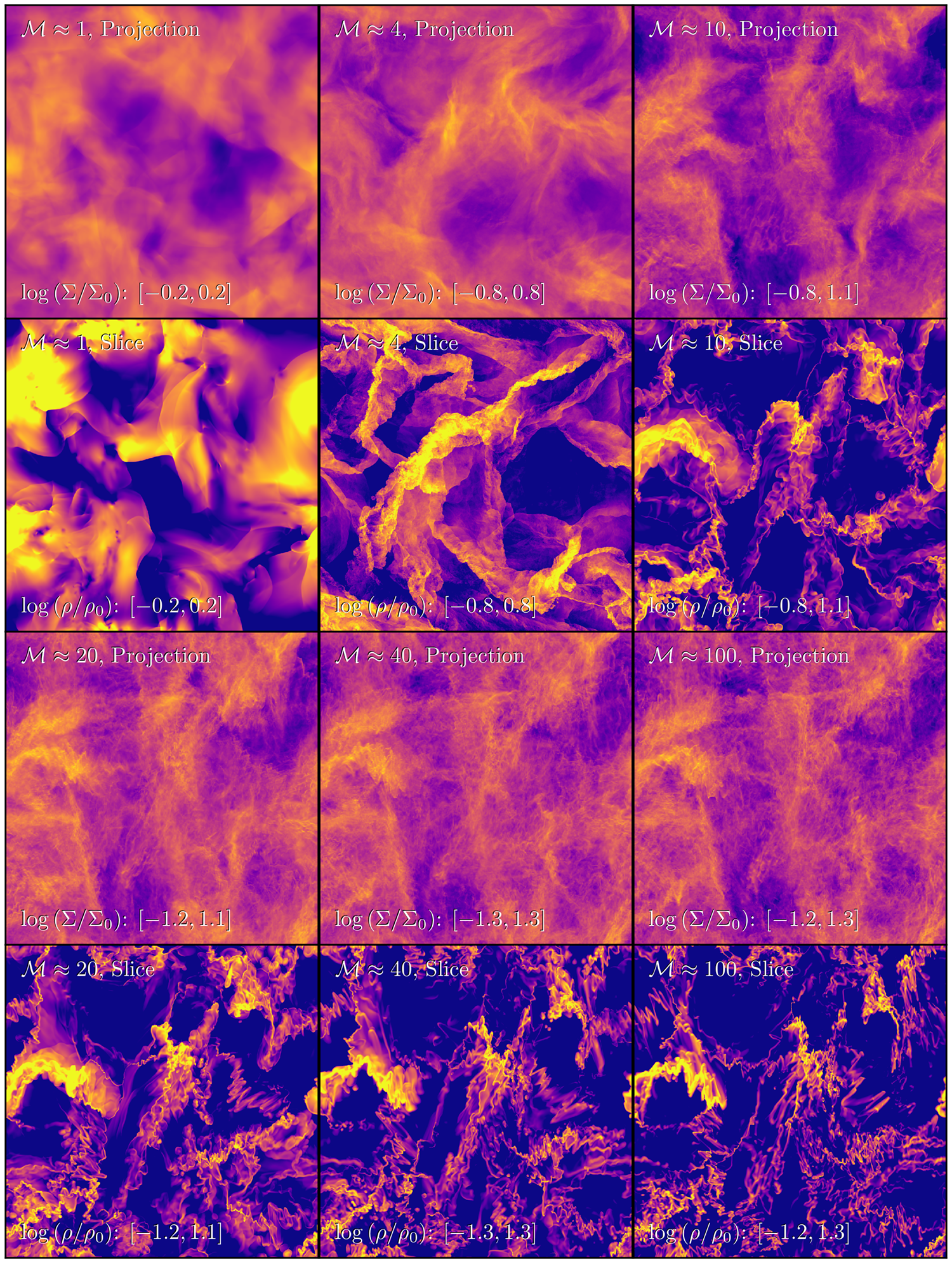}
\caption{Examples of the xy projections and xy slices from the six simulations that we use to map the projected fractal dimension, $\mathcal{D}_{\text{p}}$, to the 3D fractal dimension, $\mathcal{D}_{\text{3D}}$, with rms Mach number, $\mathcal{M} = 1$, $\mathcal{M} = 4$, $\mathcal{M} = 10$, $\mathcal{M} = 20$, $\mathcal{M} = 40$ and $\mathcal{M} = 100$, all shown at $t \approx 2 \, T$. The projections are shown in units of $\log(\Sigma/\Sigma_0)$, where $\Sigma_0$ is the mean column density and the slices are shown in units of $\log(\rho/\rho_0)$, where $\rho_0$ is the mean density. The upper and lower limits (from dark blue to yellow) of the densities are shown in the lower-left of each plot. An animation of this figure is available in the online version.}
\label{fig:12panel}
\end{figure*}

\section{Turbulence and Fractal Theory} \label{sec:theory}

\subsection{Turbulence}
Turbulence is a fluid instability that occurs when the Reynolds number $(\text{Re} \sim LV/\nu )$ becomes large, for a system of characteristic size $L$, with a velocity $V$ and kinematic viscosity $\nu$. For subsonic flows with high Reynolds number \cite{Kolmogorov1941} used dimensional arguments to show that there exists a power-law scaling, $k^{-5/3}$, where $k = 2\pi/\ell$ is the wavenumber and $\ell$ a length, that uniquely defines how the energy is moved between particular length scales (the slope of the power spectrum in the energy cascade) in the turbulence. Similar power spectrum models were used to understand supersonic turbulence. This is most relevant for the ISM as the determination of the power spectrum for observed MCs has led us to realise that MCs are subject to supersonic turbulent dynamics \citep{Burgers1948,Larson1981,Solomon1987,Ossenkopf2002,Elmegreen2004,Heyer2004,MacLow2004,Krumholz2005,Ballesteros2007,RomanDuval2011,Federrath2013}. 

Turbulence can generally be characterised by three important scales: (1) the driving scale, $\ell_\text{D}$, is the length scale where the forces that are driving the fluid occur, (2) the dissipation scale, $\ell_\nu$, is the length scale where dissipative effects turn bulk motion into random thermal motion and finally, (3) the inertial range, $\ell_\nu < \ell < \ell_\text{D}$, where energy is transported from the large, driving scale until dissipation effects take over. For turbulence driven at supersonic velocities (that is $\mathcal{M} > 1$ at scale $\ell_\text{D}$) the transport mechanism for the energy through the scaling range is dominated by shocks \citep{Burgers1948}. In contrast, in subsonic turbulence (driven at $\mathcal{M} < 1$) the energy is transported through 3D vortices \citep{Kolmogorov1941,Frisch1995}. Time scales in the turbulence are conventionally measured in units $T$, the turbulence crossing time, defined as $T = L/V$. 

\subsection{Fractal Geometry and the Fractal Dimension}
Molecular clouds are frequently inferred to be fractals. Observations of CO lines from molecular clouds show that they are organised into fractal structures, with substructures of clouds being continuously resolved, even at the highest spatial resolution achievable \citep{Falgarone1996,Stutzki1998,Chappell2001,Kauffmann2010,Schneider2013,Kainulainen2014,Rathborne2015}. To characterise a fractal, we calculate its fractal dimension. 

The fractal dimension, $\mathcal{D}$, of an object gives us information about the space-filling properties based upon the geometric structure \citep{Burrough1981,Paladin1987}. For example, if we have a cloud that is distributed diffusely throughout 3D space then $\mathcal{D}$ of the density field will be 3, indicating that the cloud fills all of the 3D space. However, if the cloud is organised into planar shocks and filaments, as they are for supersonic, star-forming MCs \citep{Burgers1948,Federrath2012,Federrath2016a,Andre2017,Federrath2018,Mocz2018}, then we should expect $\mathcal{D}$ to be between 2 (the dimension of a planar shock) and 1 (the dimension of a filament). Previous studies have related $\mathcal{D}$ to important statistics of the underlying turbulent dynamics, as well as the general structure of the clouds \citep{Stutzki1998,Chappell2001,Elmegreen2002,Sanchez2005,Elia2007,Kowal2007,Elmgreen2009,Federrath2009,Konstandin2012,Konstandin2015,Kritsuk2013,Sava2016}.

One must be careful, however, because there have been a diverse range of fractal dimensions that have been calculated for molecular clouds \citep{Elmegreen1996,Stutzki1998,Chappell2001,Elmegreen2004,Sanchez2005,Federrath2009}, and each of which need not correspond to the same $\mathcal{D}$. Furthermore fractal dimensions are degenerate. This means completely different geometries can produce the same $\mathcal{D}$ (see the Appendix in \citealt{Paladin1987} for a visualisation).

In this study we explore the mass-length dimension, $\mathcal{D}$, which comes from the scaling $m(\ell) \propto \ell^\mathcal{D}$, where $m$ is the mass in the cloud at length scale $\ell$. We explore the scale dependence of $\mathcal{D}$, which provides us information about the hierarchical structure of the clouds \citep{Paladin1987,Stutzki1998,Chappell2001,Kritsuk2013}. If there is no scale dependence it means that each hierarchy in the cloud has the same universal organisation. This is called a monofractal, whereby a single $\mathcal{D}$ is sufficient to describe the way the quantities scale (for all length scales, for example). If $\mathcal{D}$ changes with length scale, however, the properties of the hierarchy change too, meaning that the clouds are made up of a set of fractals. This kind of object is called a multifractal. It is used to describe multiplicative cascade processes and has been previously proposed for models of the density hierarchies in MCs \citep{VonWeizsacker1951,Paladin1987,Fleck1996,Stutzki1998,Chappell2001,Kowal2007}. \\


\section{Simulations} \label{sec:sims}

\subsection{Turbulent Molecular Cloud Model}
In this study we use six purely hydrodynamical simulations of molecular clouds to explore the scale dependence of the mass-length fractal dimension. For each of the six simulations we solve the compressible Euler equations,
\begin{align}
\frac{\partial \rho}{\partial t} + \nabla \cdot (\rho \mathbf{v}) &= 0, \\ 
\frac{\partial \mathbf{v}}{\partial t} + (\mathbf{v} \cdot \nabla)\mathbf{v} &= - \frac{1}{\rho} \nabla P + \mathbf{F},
\end{align}
where $\rho$ is the density, $\mathbf{v}$ the velocity, $P$ the pressure, following an isothermal equation of state, $P = c_s^2 \rho$, where $c_s$ is the speed of sound, and $\mathbf{F}$ the forcing function that drives the turbulence through a mixture of solenoidal $(\nabla \cdot \textbf{F} = 0)$ and compressive $(\nabla \times \mathbf{F} = 0)$ modes. For details we refer the reader to \cite{Federrath2010}. To summarise, $\mathbf{F}$ is an Ornstein-Uhlenbeck process that satisfies a stochastic differential equation which is parameterised by an autocorrelation time-scale, $T = L/(2c_s\mathcal{M})$, where $L$ is the integral scale of the turbulence and $\mathcal{M}$ is the desired root-mean-squared (rms) Mach number of the flow. 

\begin{table}
\caption{Simulation data and parameters.}
\centering
\begin{tabular}{r@{}lccr@{}l}
\hline
\hline
\multicolumn{2}{c}{rms $\mathcal{M}$}& Native Simulation & Number of & \multicolumn{2}{c}{Time}\\
\multicolumn{2}{c}{$(\pm1\sigma)$} & Grid Resolution & Time Slices & \multicolumn{2}{c}{Interval} \\
\hline
1.01\, & $ \pm$ 0.05 & $1024^3$ & 71 & 2 $\leq t/T$&$\leq$ 9 \\
4.1\, & $ \pm$ 0.2 & $10048^3$ & 71 & 2 $\leq t/T$&$\leq$ 9  \\
10.2\, & $ \pm$ 0.5 & $1024^3$ & 71 & 2 $\leq t/T$&$\leq$ 9\\
20\, & $ \pm$ 1 & $1024^3$ & 71 & 2 $\leq t/T$&$\leq$ 9 \\
40\, & $ \pm$ 2 & $1024^3$ & 71 & 2 $\leq t/T$&$\leq$ 9 \\
100\, & $ \pm$ 5  & $1024^3$ & 71 & 2 $\leq t/T$&$\leq$ 9 \\
\hline
\hline
\end{tabular} \\
\begin{tablenotes}
\item{\textit{\textbf{Notes:}} Column (1): the rms turbulent Mach number of the simulation $\pm$ the 1$\sigma$ temporal fluctuations. Column (2): the native 3D grid resolution of each of the simulations. Column (3): the number of time slices that we use for temporal averaging, for both the 2D slice and 2D projection data. Column (4): the time interval in units of large-scale turnover times.}
\end{tablenotes}
\label{tb:simtab}
\end{table}

We change the $\mathcal{M}$ for each of the six different simulations. The parameter set of the simulations is listed in detail in Table \ref{tb:simtab}. We explore the Mach numbers $\mathcal{M}=1,4,10,20,40$ and $100$, over seven turnover times $(7\,T)$ in the regime of fully-developed supersonic turbulence, established after $t \geq 2\,T$ \citep{Federrath2009,Price2010}. This gives us a broad overview of the behaviour of the clouds ranging from transonic, slightly compressible flows, all the way to highly supersonic, and highly compressible flows that are saturated with shocks \citep{Federrath2013}.


We solve the Euler equations in a cube with periodic boundaries. We discretise the cubic domain into $1024^3$ grid elements for the $\mathcal{M} = 1,10,20,40,100$ simulations, and $10048^3$ grid elements for the $\mathcal{M}=4$ simulation. The $10048^3$ simulation was run on 65536 compute cores, corresponding to $\approx$ 50 million compute hours, on SuperMUC, a supercomputer at the Leibniz Supercomputing Centre in Munich, Germany. This simulation is the largest supersonic turbulence simulation to date. For more details see \cite{Federrath2016a} and \cite{Federrath2018}. We use the same initial conditions for all simulations: a homogeneous medium at rest with $\rho(x,y,z,t=0) = \text{const.}$, $\mathbf{v}(x,y,z,t=0)=0$, and the same random seed for the forcing function. Hence the only difference between the simulations is the rms Mach number, $\mathcal{M}$.

\begin{figure}
\centering
\includegraphics[width=\linewidth]{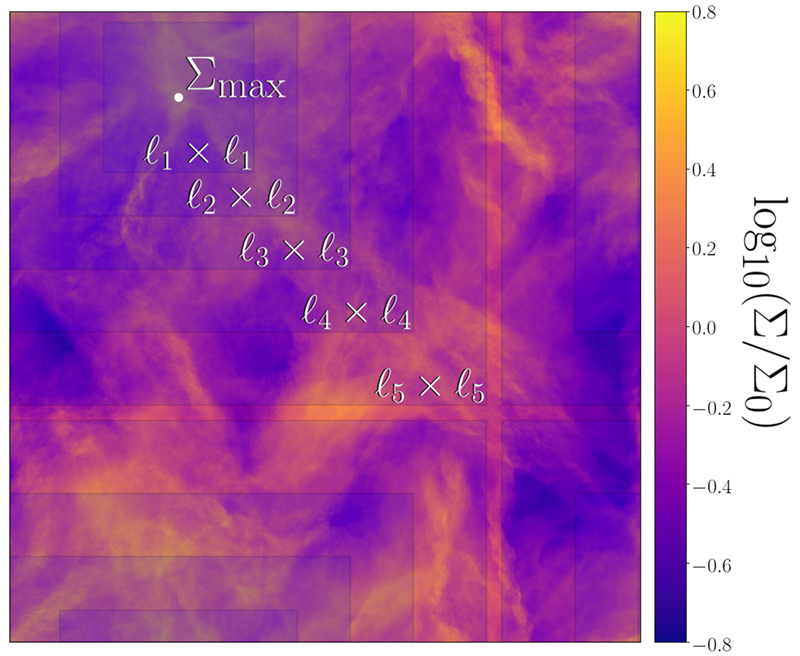}
\caption{A cartoon illustration of the mass-length method showing the expansion of an $\ell\times\ell$ square in the 2D cloud data. The square is initialised at the $\Sigma_{\text{max}}$ grid coordinate. In this cartoon, $\ell_1 < \ell_2 < \hdots < \ell_5$, represents the length scale hierarchy that we use to calculate the fractal dimension. The $\ell\times\ell$ is wrapped around the 2D data, encoding in the periodic boundary conditions of the simulation.}
\label{fig:schematic}
\end{figure}

\subsection{Data Types}
In this study we utilise two types of simulation PP data to perform our fractal dimension analysis: 2D slices through the origin in the $z=0$ plane, $\rho(x,y,0,t)$, and 2D projections of $\rho(x,y,z,t)$, the column density, integrated along the $z$-axis, 
\begin{equation} \label{eq:project}
\Sigma(x,y,t) = \int^{L/2}_{-L/2} \rho(x,y,z,t) \diff z.     
\end{equation}
Figure \ref{fig:12panel} shows a single snapshot in time $t = 2 \, T$ of the 2D projected and sliced data for each simulation.

In the absence of magnetic fields our turbulence simulations are isotropic in a statistical sense, i.e. when averaged over time, e.g., \cite{Federrath2009,Federrath2010}. This lets us perform our study only on the $xy$ projections and slices of the clouds, whilst still being representative of 2D slices through any section of the clouds, and 2D projections through any viewing angle. We use these two data types to calculate the $\mathcal{D}$ of the simulated turbulent cloud.

\begin{figure*}
\centering
\includegraphics[width=\linewidth]{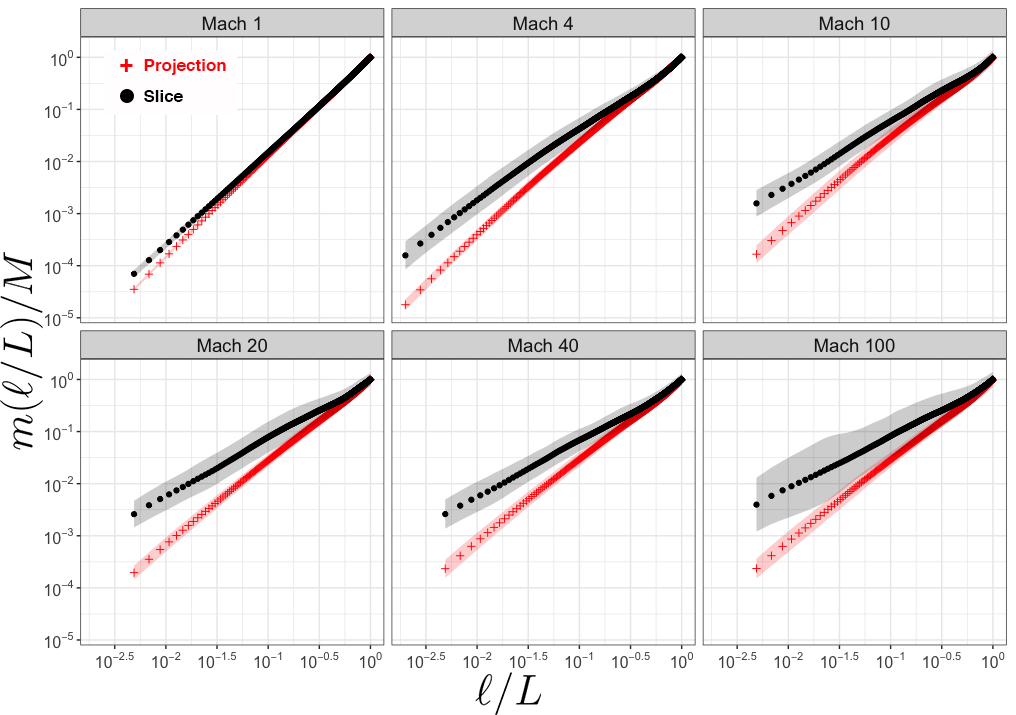}
\caption{Mass, $m/M$ as a function of length scale $\ell/L$ for each of the six different Mach numbers. The mass-length curves for projections are shown in red, with $1\sigma$ temporal fluctuations indicated by the band and likewise, for the slice data, shown in black. We see the temporal fluctuations increase with Mach number, and increase on small scales, due to the intermittency of the maximum density in the turbulent clouds.}
\label{fig:MLcurves}
\end{figure*}

\section{Mass-Length Fractal Dimension Analysis} \label{sec:MethodML}

\subsection{Mass-Length Method}

We focus on a single method for calculating the fractal dimension of the simulated, 2D cloud data --- the mass-length method\footnote{The Python 2.7 code used to implement the mass-length fractal dimension method described in this paper is available here: \url{github.com/AstroJames/FractalGeometryofPolaris}. This code supports the construction of the fractal dimension curves on periodic boundary conditions, for 2D PP simulated data, or open boundary conditions, for real PP MC data.}, with corresponding mass-length fractal dimension, $\mathcal{D}$. We will refer to this as \textit{the} fractal dimension, herein, to avoid unnecessary nomenclature. To implement the mass-length method we assume that there exists a power-law scaling between the amount of mass, $m$, within a $\ell\times\ell$ square region of the 2D projection or 2D slice of the cloud -- \mbox{either} $\Sigma(x,y,t)$ for the projected data, or $\rho(x,y,0,t)$ in the slice data. This power-law scaling of the mass and length scales has been observed in real MCs \citep{Larson1981,Myers1983,Falgarone1996,Roman-Duval2010,Donovan-Meyer2013}. The mass is given by

\begin{equation}\label{eq:ML}
m(\ell/L) \propto \left(\ell/L\right)^{\mathcal{D}},
\end{equation}
where we use the dimensionless $\ell/L$ for the length scales in the cloud. The constant of proportionality will just be $M$, the total mass, since when $\ell = L$, $m(1) \propto 1^{\mathcal{D}} = M$. We explore how the fractal dimension changes with spatial scale, i.e. we further assume that the scaling exponent (the fractal dimension) depends upon $\ell/L$,
\begin{equation}\label{eq:MLS}
m(\ell/L) = M (\ell/L)^{\mathcal{D}(\ell/L)}.
\end{equation}
We use Equation \ref{eq:MLS} to then define the fractal dimension at length scale $\ell_i/L$,
\begin{equation} \label{eq:MLS2}
    \mathcal{D}(\ell_i/L) = \frac{\log(m_i/M)}{\log(\ell_i/L)},
\end{equation}
where the $\ell_i/L$ length scale can be interpreted as all nested length scales up to the length scale $\ell_i/L$, i.e. $\ell_0/L < \ell_1/L < \ell_2/L < \hdots < \ell_i/L$, where $\ell_0/L$ is the smallest possible length scale in the cloud, and $m_i$ the corresponding mass. By doing this we impose a length scale hierarchy in the cloud, which has been done similarly, but in a different context, e.g., recently by \cite{Squire2017}. The mass-length calculation is usually performed with the limit of $\ell/L$ approaching $1$, resulting in a single $\mathcal{D}$ for an object. However, if we allow ourselves to calculate a $\mathcal{D}$ for each $\ell_i/L$, using all spatial scales below $\ell_i/L$, we can calculate a $\mathcal{D}$ for each length scale in the cloud. This treats the cloud like a nested set of density objects, each with its own $\mathcal{D}$. Next, we must choose a position in the cloud to initialise the $\ell\times\ell$ square.

\subsection{Defining the Centre for the Mass-Length Method}
To construct our $\ell\times\ell$ square we need to define coordinates that will be the centre of the square. \citet{Sanchez2005} and \citet{Federrath2009} suggest that we first identify the highest density positions in the clouds and choose these as our coordinates for the centre. We pick the simplest option and define the centre position for constructing the fractal dimension from the grid coordinate corresponding to the maximum density. We denote these as $\Sigma_{\text{max}}$, for the 2D projected data, and $\rho_{\text{max}}$, for the 2D slice data. Once we have the initialisation for coordinates of the $\ell\times\ell$ square we can expand it to measure the mass-length relation. 

\begin{figure*}
\centering
\includegraphics[width=\linewidth]{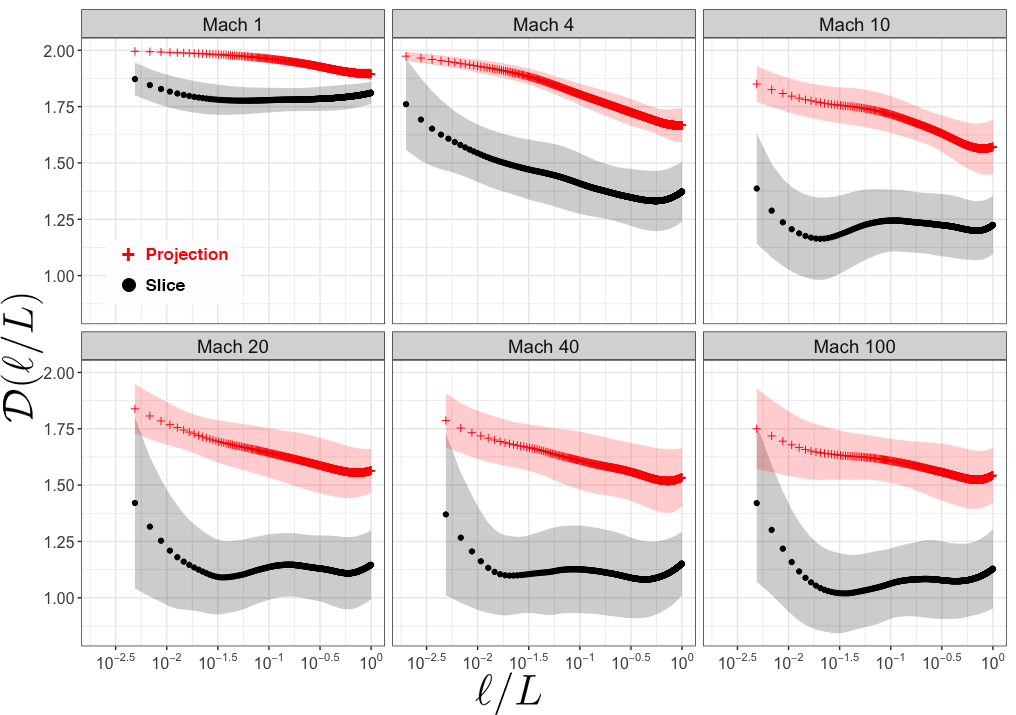}
\caption{$\mathcal{D}$ as a function of $\ell/L$ for each simulation, for both the projection and slice data, with red and grey bands around the curves showing the $1\sigma$ time fluctuations. The slice fractal dimension curves, $\mathcal{D}_{\text{s}}(\ell/L)$, (shown in black) are significantly lower in $\mathcal{D}$ than the projected fractal dimension curves, $\mathcal{D}_{\text{p}}(\ell/L)$, (shown in red) meaning that the flow is less space-filling in the slices.}
\label{fig:FDcurves}
\end{figure*}

\subsection{Constructing the Mass-Length Curves}
The $\ell\times\ell$ square is expanded from 3 grid cells (the smallest $\ell\times\ell$ square), to $N-1$, where $N$ is the number of grid cells in one dimension of the simulation (see cartoon in Figure \ref{fig:schematic}). We encode wrap-around boundary conditions from the simulation into the square expansion (please see this visualised in Figure \ref{fig:FDappendix}). The mass, $m$, is calculated for each $\ell\times\ell$ square by summing all density values multiplied by the area of each pixel within the square. We divide spatial scales by $L$, and the masses by the total mass $M$, to construct the relation in Equation \ref{eq:MLS2}. We call these curves the mass-length curves.

We build these curves for each of the 71 time slices, from $t=2\,T$ to $t = 9\,T$. Next we average over the 71 curves to produce a single curve for each simulation, using the $1\sigma$ temporal fluctuations as representative of the variation around the mean. Using these curves we can now construct the fractal dimension at each length scale in the cloud. 

\subsection{Mass-Length Curve Results} \label{sec:MLResults}
In Figure \ref{fig:MLcurves} we show the mass-length curves for the slices, illustrated in black, and the projections in red. The most important feature of these curves are the slopes, since under the $\log-\log$ transform the slopes of the curves are equal to the fractal dimension, as shown in Equation \ref{eq:MLS2}. The slope becomes shallower as the rms Mach number of the simulation increases, indicating a reduction in $\mathcal{D}$ as a function of Mach number. There are significantly different slopes in the supersonic simulations for small and large $\ell/L$. This suggests that the supersonic cloud simulations have a different mass-length scaling on large length scales compared to small length scales. The projection curves show steeper slopes, across all simulations. This means that by projecting the cloud we increase $\mathcal{D}$, resulting in smoother cloud structures in the projection. Another feature of the curves is the larger temporal fluctuations on small length scales. This is because fluctuations in $\Sigma_{\text{max}}$ and $\rho_{\text{max}}$, which are known to be intermittent \citep{Kritsuk2007,Federrath2009}, translate directly into fluctuations in the small-scale masses of the $\ell\times\ell$ square. 

From the mass-length curves we can calculate $\mathcal{D}$ at each length scale, resulting in a fractal dimension curve for each data type, and each simulation.

\begin{center}
\begin{table*}
\caption{The range of each of the twelve fractal dimension curves, shown in Figure \ref{fig:FDcurves}.}
\hfill{}
\begin{tabular}{l|cccc|cccc}
\hline
\hline
Simulation & \multicolumn{3}{c}{$\mathcal{D}_\text{p}(\ell/L)$ $\pm1\sigma$} & & \multicolumn{3}{c}{$\mathcal{D}_\text{s}(\ell/L)$ $\pm1\sigma$} & \\
Mach Number & min & & max & range & min & & max & range  \\
\hline
& & & & & & & \\
Mach 1 & $1.90 \pm 0.02 $&$ \leq \mathcal{D}_\text{p}(\ell/L) \leq $&$ 1.995 \pm 0.003$ & $0.10 \pm 0.02$  & $1.78 \pm 0.06 $&$ \leq \mathcal{D}_\text{s}(\ell/L) \leq $&$ 1.87 \pm 0.07$ & $0.1 \pm 0.1$ \\[1em]
Mach 4 & $1.66 \pm 0.08 $&$ \leq \mathcal{D}_\text{p}(\ell/L) \leq $&$ 1.97 \pm 0.02$ & $0.3 \pm 0.1$ & $1.3 \pm 0.1 $&$ \leq \mathcal{D}_\text{s}(\ell/L) \leq $&$ 1.8 \pm 0.2$ & $0.5 \pm 0.3$ \\[1em]
Mach 10 & $1.56 \pm 0.11 $&$ \leq \mathcal{D}_\text{p}(\ell/L) \leq $&$ 1.85 \pm 0.08$ & $0.3 \pm 0.2$ & $1.2 \pm 0.2 $&$ \leq \mathcal{D}_\text{s}(\ell/L) \leq $&$ 1.4 \pm 0.2$ & $0.2 \pm 0.4$ \\[1em]
Mach 20 & $1.54 \pm 0.11 $&$ \leq \mathcal{D}_\text{p}(\ell/L) \leq $&$ 1.84 \pm 0.10$ & $0.3 \pm 0.2$ & $1.1 \pm 0.2 $&$ \leq \mathcal{D}_\text{s}(\ell/L) \leq $&$ 1.4 \pm 0.4$ & $0.4 \pm 0.6$ \\[1em]
Mach 40 & $1.51 \pm 0.13 $&$ \leq \mathcal{D}_\text{p}(\ell/L) \leq $&$ 1.81 \pm 0.11$ & $0.3 \pm 0.2$ & $1.1 \pm 0.2 $&$ \leq \mathcal{D}_\text{s}(\ell/L) \leq $&$ 1.4 \pm 0.4$ & $0.4 \pm 0.6$ \\[1em]
Mach 100 & $1.52 \pm 0.13 $&$ \leq \mathcal{D}_\text{p}(\ell/L) \leq $&$ 1.75 \pm 0.16$ & $0.2 \pm 0.3$ & $1.0 \pm 0.2 $&$ \leq \mathcal{D}_\text{s}(\ell/L) \leq $&$ 1.4 \pm 0.3$ & $0.4 \pm 0.5$ \\[1em]
\hline
\hline
\end{tabular}
\hfill{}
\begin{tablenotes}
\item\textit{\textbf{Notes:}} Column (1): the Mach number of the simulation. Column (2): the minimum, maximum and range of the projected fractal dimension curves with $1\sigma$ uncertainties from the temporal fluctuations. Column (3): the same as Column 2, but for the slice fractal dimension curves. 
\end{tablenotes}
\label{tb:FDRange}
\end{table*}
\end{center}

\subsection{Constructing the Fractal Dimension Curves}
After creating the mass-length curves we fit straight lines to all of the nested length scale subsets on the $\log( \ell/L)-\log (m / M)$ data. For example, the first straight line we can fit is to the data: $\left\{(3\diff{x}/L,m_3/M),(5\diff{x}/L,m_5/M)\right\}$, where $m_3/L$ is the mass in the square $3\diff{x}\times3\diff{x}$ region, etc, and $\diff{x}$ is the cell length. The second fit is performed on the data: $\left\{(3\diff{x}/L,m_3/M),(5\diff{x}/L,m_5/L),(7\diff{x}/L,m_7/M)\right\}$, each fit always containing the data from the previous fit, encoding how the regions are nested within each other. We use the slopes of these fits to calculate a $\mathcal{D}(\ell/L)$ using Equation \ref{eq:MLS2}. For example, the slope of the fit on the second set of data, as indicated above, would be $\mathcal{D}(7\diff{x}/L)$, or the fractal dimension on spatial scale $7\diff{x}/L$. We continue this process for all $\ell/L$ up to $N-1$, always fitting on all data $\ell/L \leq \ell_i/L$ to construct $\mathcal{D}$ at each $\ell_i/L$. We call this the fractal dimension curve, herein. 

We calculate the fractal dimension curves, $\mathcal{D}_\text{s}(\ell/L)$, where the s subscript stands for slice, for 71 slices and $\mathcal{D}_\text{p}(\ell/L)$, where the p subscript stands for projection, for the 71 projections, through turnover times $t=2\,T$ to $t = 9\,T$, for each simulation. We average over the time interval to calculate a single curve both for the $\Sigma(x,y)$ and $\rho(x,y,0)$ slice data. We use the $1\sigma$ fluctuations through the time interval to construct the variation for all of the $\mathcal{D}(\ell/L)$. \\

\subsection{Fractal Dimension Curve Results} \label{sec:FDCurveResults}
In Figure \ref{fig:FDcurves} we see both, $\mathcal{D}_{\text{p}}$ and $\mathcal{D}_{\text{s}}$, as a function of $\ell/L$. $\mathcal{D}_{\text{p}}(\ell/L)$ is higher in fractal dimension than $\mathcal{D}_{\text{s}}(\ell/L)$, for all spatial scales in the turbulent clouds, as is expected from the results in \S \ref{sec:MLResults}. This means that the flow is, on all scales, more space-filling in the projections than in the slices. This is because in $\Sigma(x,y)$ we integrate across many line-of-sight slices, each of which is not exactly the same (even though they have the same statistical characteristics). Furthermore, the line-of-sight integration causes density structures to be introduced on all scales in $\Sigma(x,y)$. This decreases the contrast between the sharp, steep shocks and increases the value of the fractal dimension. We also see $\mathcal{D}_{\text{s}}$ growing between $\ell/L \approx 10^{-1.5}$ and $\ell/L \approx 10^{-1}$. This is due to the size of the voids in the slices compared to projections --- we see initial growth when the $\ell\times\ell$ square encompasses a shock, on small length scales, but then the density falls off sharply once the $\ell\times\ell$, expands over the void. This is why $\mathcal{D}_{\text{s}}$ gets pulled down for length scales larger than $\approx 1/6$ of the integral scale. 

In the slice curves we see the strongest spatial dependence (i.e. largest range in $\mathcal{D}$) in the $\mathcal{M}=4$ cloud, with $1.3 \pm 0.1 \leq \mathcal{D}_{\text{s}}(\ell/L) \leq 1.8 \pm 0.2$. This may be because in this model the dynamics of the cloud changes significantly, since the turbulence undergoes a transition between the subsonic and supersonic regimes \citep{Federrath2018}. We see this as a smooth transition in both the slices and projections, in the top, centre panel of Figure \ref{fig:FDcurves}. The smoothness hints that the 3D vortices (and 2D projections of vortices) in the subsonic turbulence must smoothly transform into sheets, filaments and shocks --- characteristic geometrical objects of supersonic turbulence \citep{Federrath2013,Federrath2016}. We show the range of all twelve fractal dimension curves calculated in Table \ref{tb:FDRange}.

The 2D slice fractal dimension curves in the bottom three panels of Figure \ref{fig:FDcurves}, suggest that the highly supersonic clouds are organised into networks of shocks, with $1.0 \pm 0.2 \leq \mathcal{D}(\ell/L) \leq 1.4 \pm 0.4$, as shown in Table \ref{tb:FDRange}. The curves further suggest that the shocks begin to dominate the clouds in the vicinity of $\mathcal{M} \approx 20$, and both the projections and the slice dimensions start becoming invariant under the growing spatial scales, suggesting a lower bound in $\mathcal{D}$. In contrast to the shocks found in the high $\mathcal{M}$ clouds, the subsonic clouds show almost completely space-filling density structure, corresponding to diffuse clouds. This transition from diffuse, space-filling structure to shocks motivates us to construct a composite fractal dimension curve, that tries to encompass all of the information, from all of the simulations into a common frame of reference.

\section{Composite Fractal Dimension Curves}\label{sec:MethodCompCurves}
Due to the smooth transition between space-filling structures to a network of strong shocks in the geometry of the simulated clouds we are motivated to combine all of our fractal dimension curves into a common frame of reference. This creates a single curve for the 2D projection data, and a single curve for the 2D slice data. Creating these composite curves provides insight into the minimum fractal dimension, and the relation between the true and observed fractal geometry of the clouds.

\subsection{Transforming into a Single Reference Frame} 
We transform all the $\mathcal{D}(\ell/L)$ curves into the $\mathcal{M}=4$ simulation frame. For the $\mathcal{M}=1$ model, we move it into the $\mathcal{M}=4$ frame by translating $\ell_\text{D}$ of the $\mathcal{M}=1$ model to the sonic scale in the $\mathcal{M}=4$ cloud. We refer to \cite{Federrath2018} for the measurement of the sonic scale. For the highly supersonic models, we move them into the $\mathcal{M}=4$ frame by shifting the length scales using Burgers' relation $\ell \propto \mathcal{M}^2$, which holds for the scaling range of the supersonic turbulence \citep{Burgers1948,Federrath2013}. For example, the length scales in the $\mathcal{M}=10$ simulation are shifted using
\begin{equation} \label{eq:elltransform}
    \left(\frac{\ell}{L}\right)_{4} = \left(\frac{\ell}{L}\right)_{10}\left(\frac{\mathcal{M}_{10}}{\mathcal{M}_4} \right)^2,
\end{equation}
where the 10 subscript on $\ell/L$ indicates the length scales in the $\mathcal{M} = 10$ frame, and similarly for the 4 subscript, and the $\mathcal{M}_4$ and $\mathcal{M}_{10}$ correspond to $\mathcal{M}=4$ and $\mathcal{M}=10$, respectively, to avoid ambiguity. These transformations allow us to create composite fractal dimension curves that span seven orders of magnitude in $\ell/L$ in the $\mathcal{M}=4$ frame. 

Now that we have a single curve for the slice and projection data, we want to construct a relation between the two. This relation could be used to map PP projected data to PP real data. However, we first relate our 2D slice curves to the real, 3D fractal dimension, $\mathcal{D}_{\text{3D}}$. This provides us with a pathway to construct a relation from the fractal geometry 2D PP projected cloud to real, 3D PPP cloud data. 

\begin{figure*}
\centering
\includegraphics[width=\linewidth]{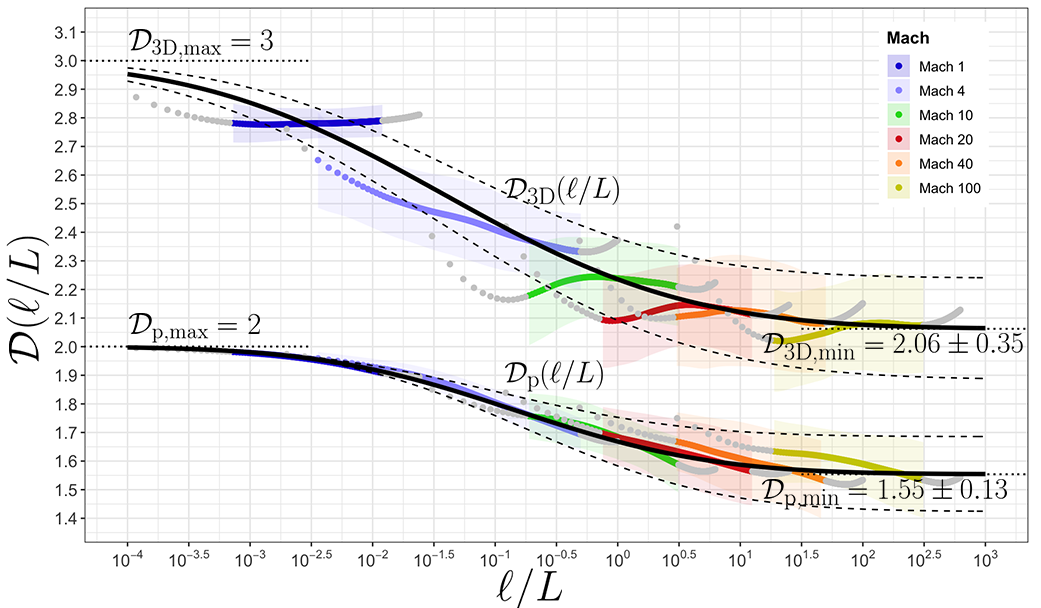}
\caption{The composite curves for $\mathcal{D}_\text{p}$ and $\mathcal{D}_{\text{3D}}$ as computed from Equation \ref{eq:st}, in the $\mathcal{M}=4$ frame, with bands around each of the curves showing the $1\sigma$ time fluctuations, coloured by Mach number. Both curves show the behaviour of the respective $\mathcal{D}$ over seven decades of spatial scales. We find that $2 \lesssim \mathcal{D}_{\text{3D}} \lesssim 3$ and $1.5 \lesssim \mathcal{D}_{\text{p}} \lesssim 2$ for all $\ell/L$. The black curves are our fit functions (Equation \ref{eq:fitfunction}), with parameters listed in Table \ref{tb:parms} and the dashed lines are the constructed errors for our fits. The fits reveal $\mathcal{D}_{\text{p,min}} = 1.55 \pm 0.13$ and $\mathcal{D}_{\text{3D,min}} = 2.06 \pm 0.35$.}
\label{fig:CompCurves}
\end{figure*}

\subsection{Mapping $\mathbf{\mathcal{D}_\text{s}}$ to $\mathbf{\mathcal{D}_{\text{3D}}}$}
We map the slice composite fractal dimension curve to the 3D fractal dimension curve utilising an analytical relation from  \mbox{\cite{Mandelbrot1983}}, 
\begin{equation} \label{eq:st}
\mathcal{D}_{\text{3D}} = \mathcal{D}_{\text{s}} + 1,
\end{equation}
which has been previously applied to turbulent flows (see \citealt{Sreenivasan1986}, for example). We apply this relation to our slice composite curve, $\mathcal{D}_s(\ell/L)$, giving us a composite curve for the 3D fractal dimension of the cloud, $\mathcal{D}_{\text{3D}}(\ell/L)$. Note that Equation \ref{eq:st} only holds for slices, but not necessarily for projections. Indeed, our final goal here is to construct a model that relates $\mathcal{D}_\text{3D}$ to $\mathcal{D}_\text{p}$.

\subsection{Composite Fractal Dimension Curve Results}

Figure \ref{fig:CompCurves} shows the composite $\mathcal{D}(\ell/L)$ for both the $\mathcal{D}_{\text{3D}}$ (mapping $\mathcal{D}_{\text{s}}$ to $\mathcal{D}_{\text{3D}}$ using Equation \ref{eq:st}) and the $\mathcal{D}_{\text{p}}$. Overall, the projected fractal dimension curves transform into the $\mathcal{M}=4$ frame well, showing a near perfect transition between the $\mathcal{M}=4$ and $\mathcal{M}=1$ curve. This is in contrast to the slice curves, which combine together better in the high $\mathcal{M}$ regimes than in the low $\mathcal{M}$ regimes.

Our composite curves suggest that both the 3D and projected clouds are organised into a multifractal with respect to the mass-length scaling. On small length scales we see that $\mathcal{D}$ for both the projected and 3D clouds is length-invariant. The mass scaling at these lengths is approximately $m \propto \ell^2$ and $m \propto \ell^{2.8}$, respectively. However, the $\mathcal{D}$ of the cloud starts changing significantly through spatial scales associated with $\mathcal{M}=1$ to $\mathcal{M}=10$, possibly from the introduction of shocks, that would reduce $\mathcal{D}$. However, we note that in the high $\mathcal{M}$ limit the rate of change in $\mathcal{D}$ slows down. Between $\mathcal{M}=40$ and $\mathcal{M}=100$ the cloud returns to a monofractal, with mass-length scaling $m \propto \ell^{1.5}$ and $m \propto \ell^{2}$, for projections and 3D cloud, respectively. 

The $\mathcal{D}$ range that our curves encompass agrees widely with fractal dimension results found in previous studies of both observations and simulations \citep{Elmegreen1996,Sanchez2005,Kowal2007,Federrath2009,Konstandin2015}. For example, \cite{Kowal2007} calculates dimensions between 2.9 and 2.5, for subsonic and supersonic magnetohydrodynamical simulations of MCs. This is similar to \cite{Federrath2009}, who measured fractal dimensions between 2.3 and 2.6 (depending upon the turbulent driving modes) and \cite{Elmegreen1996} (and references therein), who measure $\mathcal{D}$ $\approx 2.3$ for interstellar gas. Each of these findings are well encompassed by the $\mathcal{D}$ values of our curves. 

Now that we have constructed single fractal dimension curves, for both $\mathcal{D}_{\text{3D}}$ and $\mathcal{D}_{\text{p}}$ we are able to build an empirical fit using all of the $\ell/L$ data in the common frame. This is then used to measure the minimum $\mathcal{D}$ in the high $\mathcal{M}$ limit, and to comprehensively relate $\mathcal{D}_{\text{3D}}$ and $\mathcal{D}_{\text{p}}$ for all $\ell/L$. 




\subsection{Building an Empirical Fit}
We use all of the data in the common simulation frame to construct empirical fits for both the slice and projection composite curves. We choose a complementary error function, $\erfc(x) = 1 - \erf(x)$, for our fit. This is an appropriate function for modelling a smooth, asymptotic state change between two limits, mirroring the behaviour we see for our composite curve data in Figure 5. This is because the complementary error function is bound between the values,  $\lim_{x\rightarrow-\infty} [1-\erf(x)] = 2$, and $\lim_{x\rightarrow\infty} [1-\erf(x)] = 0$, and evolves monotonically between them. One such example where the $\erfc(x)$ is used in fluid dynamics is the Rayleigh problem, where a fluid is agitated by a sliding plate with constant velocity, $U$. The velocity field of the fluid becomes bound between $U$ (at the plate boundary) and 0 (infinitely far away from the plate), with a smooth, monotonic transition between them. The functional form we use is,
\begin{equation}\label{eq:fitfunction}
\mathcal{D}(\ell/L) = \frac{\mathcal{D}_{\text{max}} - \mathcal{D}_{\text{min}}}{2} \erfc \Big( \beta_1\log \left( \ell/L \right) - \beta_0 \Big) + \mathcal{D}_{\text{min}},
\end{equation}
which has four parameters: (1) $\mathcal{D}_{\text{max}}$ is the maximum possible fractal dimension, (2) $\mathcal{D}_{\text{min}}$ is the minimum fractal dimension we measure and (3) and (4), $\beta_1$ and $\beta_0$ correspond to the rate in which the error function moves from maximum to minimum $\mathcal{D}$ and the translation across the $\ell/L$ axis, respectively. We implement weighted non-linear least squares regression to fit Equation \ref{eq:fitfunction}, weighted by the $1\sigma$ temporal fluctuations on the turbulence cascades of each simulation. We estimate three of the four parameters for the slice and projection data, assuming that the $\mathcal{D}_{\text{max}} = 2$ for the $\mathcal{D}_{\text{p}}(\ell/L)$ curve, and $\mathcal{D}_{\text{max}} = 3$ for the $\mathcal{D}_{\text{3D}}(\ell/L)$ curve, as also assumed by \cite{Konstandin2015}. This is because $\mathcal{D}=3$ corresponds to completely space-filling structures in 3D space, which is what we expect in the limit of $\mathcal{M} \rightarrow 0$, corresponding to small $\ell/L$ on our curves. Likewise, for the projected dimension, we assume that the same is true but for space-filling structures on the plane (i.e. $\mathcal{D}_{\text{p,max}} = 2$).


\begin{figure*}
\centering
\includegraphics[width=\linewidth]{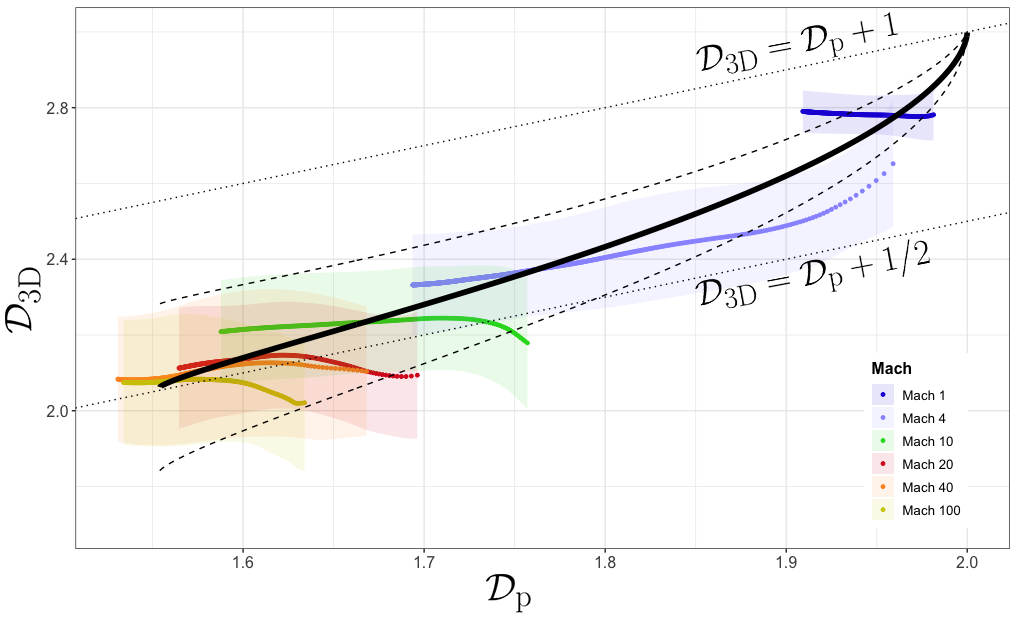}
\caption{The empirical fit (Equation \ref{eq:MappingFunction}) of the composite 3D fractal dimension curve as a function of the projected fractal dimension curve (shown in black), with Gaussian error propagation of the 1$\sigma$ temporal fluctuations shown in Figure \ref{fig:CompCurves}, indicated by the grey band. Each simulation is coloured as in Figure \ref{fig:CompCurves}. We see in the low Mach number limit (shown to the right of the plot) the relation is bounded by $\mathcal{D}_{\text{3D}} \approx \mathcal{D}_{\text{p}} + 1$ and in the high Mach number limit (shown to the left of the plot), $\mathcal{D}_{\text{3D}} \approx \mathcal{D}_{\text{p}} + 1/2$. Typical molecular clouds will have Mach numbers between these two limits, so the full model provided by Equation \ref{eq:MappingFunction} is required to describe the relation between $\mathcal{D}_\text{3D}$ and $\mathcal{D}_\text{p}$.}
\label{fig:FDMaps}
\end{figure*}

\begin{table}
\caption{The fit function parameters for Equation \ref{eq:fitfunction}, \ref{eq:MappingFunction}, and fits plotted on Figure \ref{fig:CompCurves}, estimated using weighted, non-liner least squares regression. All $1\sigma$ uncertainties are from the statistical fitting.}
\centering
\begin{tabular}{lcccc}
\hline
\hline
Fit & $\beta_0 \pm 1\sigma$ & $\beta_1 \pm 1\sigma$ & $\mathcal{D}_{\text{min}} \pm 1\sigma$ & $\mathcal{D}_{\text{max}}$ \\ 
\hline
& & & & \\
$\mathcal{D}_{\text{p}}(\ell/L)$ &  0.46  $\pm$ 0.08 & 0.56$\pm$ 0.08 & 1.55 $\pm$ 0.13 & 2 \\[1em] 
$\mathcal{D}_{\text{3D}}(\ell/L)$ &  0.63 $\pm$ 0.08 & 0.45 $\pm$ 0.03 & 2.06 $\pm$ 0.35 & 3 \\[1em] 
\hline
\hline
\end{tabular} \\
\begin{tablenotes}
\item{\textit{\textbf{Notes:}} Column (1): the fit, either $\mathcal{D}_{\text{p}}(\ell/L)$ for the projected data or $\mathcal{D}_{\text{3D}}(\ell/L)$ for the slice data that was mapped to the 3D fractal dimension using Equation \ref{eq:st}. Column (2): the $\beta_0$ parameter value, corresponding to the translation along the $\ell/L$-axis. Column (3): the $\beta_1$ parameter value, corresponding to the rate in which the error function moves between upper and lower bounds. Column (4): the minimum fractal dimension from the fit, i.e. the fractal dimension in the high $\mathcal{M}$ limit. Column (5): the maximum fractal dimension, which was assumed for both fits.}
\end{tablenotes}
\label{tb:parms}
\end{table}

We employ our empirical fit, Equation \ref{eq:fitfunction}, to calculate the asymptotic behaviour of the $\mathcal{D}(\ell/L)$ curves for large length scales, corresponding to large Mach numbers. This corresponds to determining the $\mathcal{D}_{\text{min}}$ for both of the curves. We find that $\mathcal{D}_{\text{p,min}} = 1.55 \pm 0.13$ and $\mathcal{D}_{\text{3D,min}} = 2.06 \pm 0.35$. In the 3D geometry this corresponds to the dimension of a planar shock. This is in excellent agreement with our current understanding of what kind of structures dominate the density field in the highly supersonic turbulent regime \citep{Burgers1948,Federrath2013}. It is also in agreement with the \cite{Konstandin2015} model for the fractal dimension, $\mathcal{D} = 2 + 0.96\mathcal{M}^{-0.30}$, which is developed through the Fourier transform of the mass-length relation and gives a value of 2 in the high $\mathcal{M}$ limit. 
In the projected geometry we do not see a $\mathcal{D}_{\text{p,min}} = 1$, as one might expect. This is because of the smoothing introduced by projecting a cloud, previously mentioned in \S \ref{sec:FDCurveResults}. The values of the two other estimated parameters from the fits, $\beta_0$ and $\beta_1$, are shown in Table \ref{tb:parms}. We can now use our empirical fits to comprehensively relate $\mathcal{D}_{\text{3D}}$ with $\mathcal{D}_{\text{p}}$.

\section{The $\mathcal{D}_{\text{p}}-\mathcal{D}_{\text{3D}}$ Relation} \label{sec:relation}

\subsection{Constructing the Relation}

In Figure \ref{fig:FDMaps} we plot the fits for $\mathcal{D}_{\text{p}}(\ell/L)$ and $\mathcal{D}_{\text{3D}}(\ell/L)$  against each other, shown in black, to create an empirical relation between the two quantities. This provides us with a smooth function relating the projected fractal dimension to the 3D fractal dimension,
\begin{equation}\label{eq:MappingFunction}
\mathcal{D}_{\text{3D}}(\mathcal{D}_{\text{p}}) = \Omega_1 \erfc \Bigg[ \xi_1 \erfc^{-1}\Bigg(\frac{\mathcal{D}_{\text{p}} - \mathcal{D}_{\text{p,min}}}{\Omega_2} \Bigg) + \xi_2 \Bigg] + \mathcal{D}_{ \text{3D,min}},
\end{equation}
where $\erfc^{-1}$ is the inverse complementary error function,
\begin{align*}
\xi_1 &= \frac{\beta_{1,\text{3D}} }{ \beta_{1,\text{p}}}, \\
\xi_2 &= \beta_{0,\text{3D}} - \beta_{0,\text{p}}\hspace{1mm}\xi_1,
\end{align*}
and 
\begin{align*}
    \Omega_1 &= \frac{ \mathcal{D}_{\text{3D,max}} - \mathcal{D}_{\text{3D,min}} }{2}, \\
    \Omega_2 &= \frac{ \mathcal{D}_{\text{p,max}} - \mathcal{D}_{\text{p,min}} }{2}.
\end{align*}

The parameter values for $\Omega_1$, $\Omega_2$, $\xi_1$ and $\xi_2$ are found with $1\sigma$ uncertainties in Table \ref{tb:parms2} and likewise for $\mathcal{D}_{ \text{3D,min} }$ and $\mathcal{D}_{\text{p,min}}$ in Table \ref{tb:parms}. In the above equations we use the p and 3D subscripts to indicate $\beta$ parameters and minimum and maximum fractal dimension from the $\mathcal{D}_\text{p}(\ell/L)$ and $\mathcal{D}_\text{3D}(\ell/L)$ fits, respectively. We construct the error for this relation by taking the Euclidean sum of the 1$\sigma$ temporal fluctuations from each curve, $\sqrt{ \sigma_{\text{3D}}^2 + \sigma_{\text{p}}^2 }$, where $\sigma_{\text{3D}}$ are the 1$\sigma$ fluctuations from the $\mathcal{D}_{\text{3D}}(\ell/L)$ curve, and $\sigma_{\text{p}}$ are the 1$\sigma$ fluctuations from the $\mathcal{D}_{\text{p}}(\ell/L)$ curve.


\begin{table}
\caption{Parameters values for the $\mathcal{D}_\text{p}-\mathcal{D}_\text{3D}$ relation, shown in Equation \ref{eq:MappingFunction}, each with 1$\sigma$ uncertainties.}
\centering
\begin{tabular}{cccc}
\hline
\hline
$\Omega_1 \pm 1\sigma$ & $\Omega_2 \pm 1\sigma$ & $\xi_1 \pm 1\sigma$ & $\xi_2 \pm 1\sigma$ \\
\hline
& & & \\
0.47 $\pm$ 0.18 &  0.22  $\pm$ 0.07 & 0.80 $\pm$ 0.18 & 0.26 $\pm$ 0.19 \\[1em] 
\hline
\hline
\end{tabular} \\
\begin{tablenotes}
\item{\textit{\textbf{Notes:}} Column (1): the $\Omega_1$ parameter, equal to $(1/2)(\mathcal{D}_{\text{3D,max}} - \mathcal{D}_{\text{3D,min}})$. Column (2): the $\Omega_2$ parameter, equal to $(1/2)(\mathcal{D}_{\text{p,max}} - \mathcal{D}_{\text{p,min}})$. Column (3): the $\xi_1$ parameter, equal to $\beta_{1,\text{3D}}/\beta_{1,\text{p}}$. Column (4): the $\xi_2$ parameter, equal to $\beta_{0,\text{3D}} - \beta_{0,\text{p}}\xi_1 $.}
\end{tablenotes}
\label{tb:parms2}
\end{table}

\subsection{$\mathcal{D}_{\text{p}}-\mathcal{D}_{\text{3D}}$ Relation Results}

In Figure \ref{fig:FDMaps} we plot $\mathcal{D}_{\text{3D}}(\mathcal{D}_{\text{p}})$, with the grey band showing the constructed 1$\sigma$ uncertainty. As discussed in \cite{Stutzki1998}, \cite{Sanchez2005} and \cite{Federrath2009}, since \cite{Mandelbrot1983}'s slice relation for $\mathcal{D}_{\text{s}}$ and $\mathcal{D}_{\text{3D}}$ is so simple (Equation \ref{eq:st}), one might hope that a similar linear mapping holds for the $\mathcal{D}_{\text{p}}$ to $\mathcal{D}_{\text{3D}}$ relation. Unfortunately, Figure \ref{fig:FDMaps} shows that this is not the case. The process of projecting the 3D structure of the cloud onto a 2D plane does not change the structure uniformly across all of the spatial scales involved. Instead, we find larger smoothing on small length scales than large length scales. In the low and high Mach limit (small and large spatial scales, respectively) we see the effect of projecting the geometry on the fractal dimension is bounded by a set of simple, linear relationships. In the low Mach limit $(\mathcal{M} \ll 1)$ we find that $\mathcal{D}_{\text{3D}} \approx \mathcal{D}_{\text{p}} + 1$ bounds the relation. This is by construction, because we assume that $\mathcal{D}_{\text{p,max}} = 2$ and $\mathcal{D}_{\text{3D,max}} = 3$. In the high Mach limit $(\mathcal{M} > 20)$, the relation is bounded by $\mathcal{D}_{\text{3D}} \approx \mathcal{D}_{\text{p}} + 1/2$. Molecular clouds in the Milky Way typically have turbulent Mach numbers between $\mathcal{M}=5-20$ \citep{Larson1981,Heyer2004,Federrath2013}, so the full relation, shown in Equation \ref{eq:MappingFunction} will be needed to relate the 2D projected fractal dimension to the 3D fractal dimension for real MCs.

\section{Summary and Key Findings}\label{sec:conclusions}
In this study we define a new method for calculating the mass $(m)$ - length $(\ell)$ fractal dimension, $\mathcal{D}$, where $m \propto \ell^{\mathcal{D}}$, with the aim of relating 2D projected PP (position-position) structures to the structures found in 3D PPP (position-position-position) molecular clouds. We implement our method on both 2D slices, with fractal dimension $\mathcal{D}_{\text{s}}$, and 2D projections, with fractal dimension $\mathcal{D}_{\text{p}}$ of simulated, 3D molecular clouds undergoing turbulent driving between $\mathcal{M}=1$ and $\mathcal{M}=100$. We construct curves from the fractal dimension by plotting $\mathcal{D}(\ell/L)$, where $L$ is the integral scale of the cloud, for both the slice, $\mathcal{D}_{\text{s}}(\ell/L)$, and projection, $\mathcal{D}_{\text{p}}(\ell/L)$, data. Next, we put all of the fractal dimension curves in a common frame, creating composite $\mathcal{D}(\ell/L)$ curves, across two orders of magnitude in $\mathcal{M}$ and seven orders of magnitude in $\ell/L$. We use the \cite{Mandelbrot1983} relation to map the 2D slice curve to the 3D fractal dimension curve, $\mathcal{D_{\text{3D}}}(\ell/L)$. We provide an empirical fit for the composite curves, Equation \ref{eq:fitfunction}, for both the slice and projection data, over this entire range of scales and Mach numbers. We estimate three parameters for both the slice and projection fits using weighted non-linear least squares regression, including the minimum possible $\mathcal{D}$ for a turbulent molecular cloud, Table \ref{tb:parms}. Finally, we plot our fits against each other to find a new relation between $\mathcal{D_{\text{3D}}}$ and $\mathcal{D}_{\text{p}}$, thus providing a map from the 2D projected fractal dimension to the 3D fractal dimension of the clouds. We summarise our key findings below:

\begin{itemize}
    \item We present a new method for calculating the mass $(m)$ - length $(\ell)$ fractal dimension, from the scaling relation, $m \propto \ell^{\mathcal{D}(\ell)}$, of molecular clouds. We provide a Python 2.7 code that works on 2D cloud data, with periodic boundaries or open boundaries. It can be downloaded at: \url{github.com/AstroJames/FractalGeometryofPolaris}. \\
    
    \item With our new method, we find that $\mathcal{D}_{\text{s}}(\ell/L) < \mathcal{D}_{\text{p}}(\ell/L)$ for all $\ell/L$. This is because the line-of-sight integration in the construction of the column density smooths the density structures in $\rho(x,y,z)$, resulting in a globally higher $\mathcal{D}$, i.e. more space-filling structures. However, the smoothing from the projection does not act uniformly across all $\ell/L$, and the effect is greater at large $\ell/L$ and high $\mathcal{M}$ than on small scales. \\
    
    \item We find that both the sliced and the projected density fields exhibit a multifractal structure that goes from being independent of length at low Mach number $(\mathcal{M} < 1)$, to length dependent through $\mathcal{M}=1$ to $\mathcal{M}=10$, and back to length independence at very high Mach number $(\mathcal{M} \gg 10)$. \\
    
    \item We construct composite curves utilising the sonic scale and \cite{Burgers1948} relation in the supersonic cascade and combine all fractal curves into the same reference frame. Using our composite curves we find that the geometry of the projected density field transforms smoothly from subsonic, space-filling structures, to highly supersonic shock-filled structures. \\
    
    \item The fractal dimension values encompassed in our composite curves agree with previous studies of the fractal dimensions of simulations and observations of molecular clouds, which find $\mathcal{D} \approx 2.0 - 2.7$ \citep{Elmegreen1996,Sanchez2005,Kowal2007,Federrath2009,Konstandin2015}. \\

    \item We find a new relation between the 2D projected (PP data) and 3D (PPP data) fractal dimension of turbulent clouds, $$\mathcal{D}_{\text{3D}}(\mathcal{D}_{\text{p}}) = \Omega_1 \erfc \Bigg[ \xi_1 \erfc^{-1}\Bigg(\frac{\mathcal{D}_{\text{p}} - \mathcal{D}_{\text{p,min}}}{\Omega_2} \Bigg) + \xi_2 \Bigg] + \mathcal{D}_{ \text{3D,min}},$$ shown in Equation \ref{eq:MappingFunction}, where the minimum 3D fractal dimension, $\mathcal{D}_{\text{3D,min}} = 2.06 \pm 0.35$, the minimum projected fractal dimension, $\mathcal{D}_{\text{p,min}} = 1.55 \pm 0.13$, $\Omega_1 = 0.47 \pm 0.18$, $\Omega_2 = 0.22 \pm 0.07$, and $\xi_1$ and $\xi_2$ are $0.80 \pm 0.18$ and $0.26 \pm 0.19$, respectively. The measured $\mathcal{D}_{\text{3D,min}}$ agrees with the theoretical minimum in \cite{Konstandin2015} model, $\mathcal{D} = 2 + 0.96\mathcal{M}^{-0.30}$. This corresponds to a cloud saturated with strong shocks in the high $\mathcal{M}$ limit. We observe that this condition is sufficiently met at $\mathcal{M} > 20$. In the low and high $\mathcal{M}$ limits our relation is bounded by $\mathcal{D}_{\text{3D}} \approx \mathcal{D}_{\text{p}} + 1$ and $\mathcal{D}_{\text{3D}} \approx \mathcal{D}_{\text{p}} + 1/2$, respectively. \\
    
    \item The new relation, coupled with our method for calculating the fractal dimension as a function of length scale, could be used to probe the 3D structure (and other quantities) of molecular clouds, based purely on the observed, 2D projected cloud. 
\end{itemize}

\section*{Acknowledgements}\label{sec:acknowledgments}
J.~R.~B.~thanks all those who helped with reading and discussing this piece of science, especially Mairead MacKinnon. C.~F.~acknowledges funding provided by the Australian Research Council (Discovery Project DP170100603, and Future Fellowship FT180100495), and the Australia-Germany Joint Research Cooperation Scheme (UA-DAAD). R.~S.~K. acknowledges support from the Deutsche Forschungsgemeinschaft via SFB 881, ``The Milky Way System" (sub-projects B1, B2 and B8). We further acknowledge high-performance computing resources provided by the Leibniz Rechenzentrum and the Gauss Centre for Supercomputing (grants~pr32lo, pr48pi and GCS Large-scale project~10391), the Partnership for Advanced Computing in Europe (PRACE grant pr89mu), the Australian National Computational Infrastructure (grant~ek9), and the Pawsey Supercomputing Centre with funding from the Australian Government and the Government of Western Australia, in the framework of the National Computational Merit Allocation Scheme and the ANU Allocation Scheme. The simulation software FLASH was in part developed by the DOE-supported Flash Centre for Computational Science at the University of Chicago.

\bibliographystyle{mnras.bst}
\bibliography{bibl.bib}

\appendix
\section{Selected Mass-Length Method Animation Frames}

\begin{figure*}
\centering
\includegraphics[width=\linewidth]{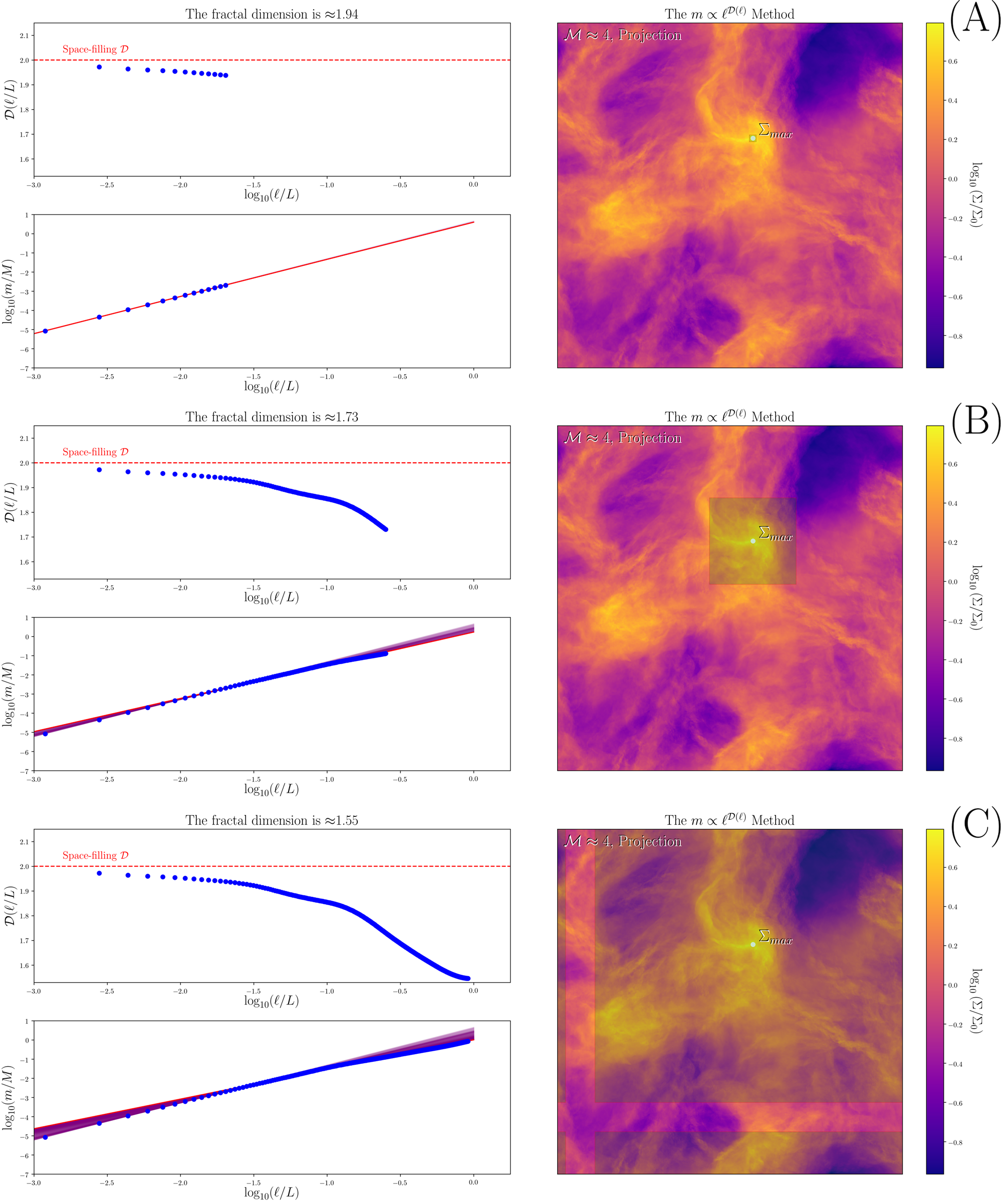}
\caption{Three snapshots of the mass-length method at different $\ell\times\ell$ (shown as the transparent blue square) on a single time step of the projected cloud data. The top-left plot is the fractal dimension as a function of length scale (the fractal dimension curves, shown in Figure \ref{fig:FDcurves}). The bottom-left plot is the mass as a function of length scale (the mass-length curves, shown in Figure \ref{fig:MLcurves}). Each step in $\ell$ shows a new line being fit to the data. Previous fits are retained (shown as a purple shadow) to visualise the change in the slope as the $\ell\times\ell$ region expands. The slope of the line corresponds to the fractal dimension at that length scale. (A) is early on in the implementation, with small $\ell\times\ell$ and (B) and (C) show the middle and towards the end of the implementation. A full animation of this figure is available in the online version.}
\label{fig:FDappendix}
\end{figure*}



\label{lastpage}

\newpage

\end{document}